\documentclass[%
reprint,
onecolumn,
superscriptaddress,
nofootinbib,
 amsmath,amssymb,
 aip,
 jmp,
]{revtex4-2}

\usepackage{graphicx}% Include figure files
\usepackage[hidelinks]{hyperref}
\usepackage{float}
\usepackage{booktabs}
\usepackage{textcase}
\usepackage{makecell} % line breaks within table cells

\usepackage[locale=US,mode=math]{siunitx}
\DeclareSIUnit\liter{l}
\DeclareSIUnit\bar{bar}
\usepackage[capitalize]{cleveref} 
\usepackage[version=4]{mhchem}
\usepackage{xfrac}
\usepackage{subcaption}
\newcommand{\uxodt}{µXODT}
\DeclareSIUnit\liter{l}	 
\newcommand{\lifetime}{$\tau=\SI[ round-mode=uncertainty, round-precision=2, separate-uncertainty=true]{35.62 \pm 1.66}{\second}$}
	 
\begin{document}

\title{Micro-integrated crossed-beam optical dipole trap system with long-term alignment stability for mobile atomic quantum technologies}

\author{Marc Christ}
 \email{marc.christ@fbh-berlin.de}
 \affiliation{Ferdinand-Braun-Institut (FBH), Gustav-Kirchhoff-Straße 4, 12489 Berlin, Germany}
  \affiliation{Humboldt-Universit\"at zu Berlin, Institut f\"ur Physik \& IRIS Adlershof, Newtonstraße 15, 12489 Berlin, Germany}
\author{Oliver Anton}
 \affiliation{Humboldt-Universit\"at zu Berlin, Institut f\"ur Physik \& IRIS Adlershof, Newtonstraße 15, 12489 Berlin, Germany}
\author{Conrad Zimmermann}
 \affiliation{Ferdinand-Braun-Institut (FBH), Gustav-Kirchhoff-Straße 4, 12489 Berlin, Germany}
\author{Victoria A Henderson}
 \affiliation{Humboldt-Universit\"at zu Berlin, Institut f\"ur Physik \& IRIS Adlershof, Newtonstraße 15, 12489 Berlin, Germany}
 \affiliation{Now at RAL Space, Science and Technology Facilities Council, Rutherford Appleton Laboratory, Harwell, Didcot, Oxfordshire, OX11 0QX, UK}
\author{Elisa Da Ros}
 \affiliation{Humboldt-Universit\"at zu Berlin, Institut f\"ur Physik \& IRIS Adlershof, Newtonstraße 15, 12489 Berlin, Germany}
\author{Markus Krutzik}
 \affiliation{Ferdinand-Braun-Institut (FBH), Gustav-Kirchhoff-Straße 4, 12489 Berlin, Germany}
 \affiliation{Humboldt-Universit\"at zu Berlin, Institut f\"ur Physik \& IRIS Adlershof, Newtonstraße 15, 12489 Berlin, Germany}

\begin{abstract}
Quantum technologies extensively use laser light for state preparation, manipulation, and readout. For field applications, these systems must be robust and compact, driving the need for miniaturized and highly stable optical setups and system integration. In this work, we present a micro-integrated crossed-beam optical dipole trap setup, the \uxodt, designed for trapping and cooling \ce{^{87}Rb}. This fiber-coupled setup operates at \SI{1064}{\nano\meter} wavelength with up to \SI{2.5}{\watt} optical power and realizes a free-space crossed beam geometry. The \uxodt~precisely overlaps two focused beams ($w_0\approx\SI{33}{\micro\meter}$) at their waists in a \SI{45}{\degree} crossing angle, achieving a position difference \SI{\leq 3.4}{\micro\meter} and \num{0.998} power ratio between both beams with long-term stability. We describe the design and assembly process in detail, along with optical and thermal tests with temperatures of up to \SI{65}{\celsius}. The system's volume of \SI{25}{\milli\liter} represents a reduction of more than two orders of magnitude compared to typically used macroscopic setups, while demonstrating exceptional mechanical robustness and thermal stability. The \uxodt~is integrated with a \ce{^{87}Rb} 3D MOT setup, trapping \num{3E5} atoms from a laser-cooled atomic cloud, and has shown no signs of degradation after two years of operation.
\end{abstract}

\maketitle

\section{Miniaturized optical systems for compact quantum technology}

Quantum technologies are transitioning from research in dedicated laboratory environments to practical applications for field deployment. A common characteristic of a wide range of quantum technologies is the use of laser light for state preparation, manipulation, and readout. This is particularly relevant for atomic quantum sensors, which can detect various external fields or forces, or act as frequency reference with a significant improvement in sensitivity and precision compared to classical counterparts \cite{Kitching2011,Bongs2019,Canuel2020, Peters1999}. Deployment on mobile platforms on ground or in space opens various applications in an extended parameter regime \cite{Becker2018,Frye2021,Leveque2022,Abend2023}, but demands a high degree of system miniaturization, integration, and robustness against environmental conditions. This promotes the need for compact and highly stable optical setups without the need for realignment, also in the physics package, the sensor's subsystem containing the atomic quantum system.

Many atomic quantum sensors require (ultra-)cold atoms, and thus magneto-optical traps (MOTs) are utilized to initially trap and further cool the atoms within a vacuum chamber \cite{Foot1991,Perrin2009}. Multiple approaches are under development for miniaturization of the physics package \cite{Rushton2014}, including grating-based magneto-optical traps (gMOT) \cite{Vangeleyn2010, Cotter2016, Chen2022, Heine2024, Calviac2024}, additive manufacturing \cite{Madkhaly2021, Cooper2021, Christ2024}, and atom chips \cite{Keil2016, Kassner2019}. These approaches result in compact vacuum setups and optical systems \cite{Lee2022,Straatsma2015}, which are also suitable for space-based platforms\cite{Lachmann2021,Elliott2023}. Optical dipole traps (ODT) are widely used for cold-atom generation and manipulation\cite{GRIMM200095, Bartenstein2004, Arnold2011, Navon2021a}. They are used for lensing ultra-cold atoms \cite{Kanthak2021,Albers2022} and are essential for dual species atom interferometry \cite{Schuldt2015}. While a crossed-beam ODT configuration significantly increases the atom confinement compared to a single beam ODT, it demands precise and stable alignment of two focused beams. For mobile applications with mechanical and thermal environmental loads, this requires active realignment or even pointing stabilization in the currently established setups based on macroscopic optical components \cite{Vogt2020}. This is often impractical and would increase the system complexity and overall size, weight, and power consumption (SWaP) of the physics package.

One approach at FBH that leads to a substantial miniaturization is micro-integration, which is an established technique, e.g., for diode laser modules \cite{Schiemangk2015,Wicht2017,Kuerbis2020}. Here, multiple components such as laser diodes, electro-optical elements, and optical components are integrated on a functionalized unified substrate, the micro-optical bench, often using adhesive bonding\cite{Christ2023}. This yields a significant reduction of system size while achieving free-space optical setups with high mechanical stability, thus resulting in an optical system without the need for realignment or included opto-mechanical components. This approach demonstrated its stability in various drop tower experiments and sounding rocket missions \cite{Schkolnik2016, Dinkelaker2016, Doeringshoff2019}. 
 
In this work, a micro-integrated optical system is used for cold-atom manipulation within a physics package. A crossed-beam optical dipole trap system, the \uxodt, is realized. Such a system benefits strongly from miniaturization and stability of micro-integration. The design, assembly, and qualification are discussed and results from the implementation in a cold-atom system are presented.

\section{Micro-integrated, crossed-beam optical dipole trap}
\label{sec:uxodt_design_assembly_tests}
The \uxodt~is designed for integration into a cold-atom system (Infleqtion RuBECi), which is similar to a system deployed on the ISS \cite{Elliott2023, Williams2024}, but not equipped with an atom chip. It features a glass two-cell vacuum system (cross section \qtyproduct[product-units=power]{26 x 26}{\milli\metre}) for the 3D MOT and the 2D+ MOT, which includes dispensers of rubidium and potassium in natural abundance. Multiple optical components are positioned around the system to generate the MOT geometry, perform absorption imaging detection, and facilitate planned quantum memory experiments \cite{DaRos2022}.  The system has been used in previous works on optimized 3D MOT loading using machine learning algorithms \cite{Anton2024}. A crossed-beam ODT extends the capabilities of this setup, serving as a building block for (ultra-)cold atom generation and manipulation\cite{Perrin2009}. It is designed as far-off resonance trap (FORT) \cite{GRIMM200095} operated at $\lambda=\SI{1064}{\nano\meter}$, a common choice due to the broad availability of high-power laser sources and optical components.

To achieve a large potential depth for atom trapping with an ODT, high optical intensities, and thus narrow beam waists combined with high powers are required. A crossed-beam layout further increases the 3D atom confinement and generates a more symmetric trap, while the polarization of the beams needs to be perpendicular to each other to prevent interference in the overlap. The waist radius of an ideal Gaussian beam with input radius $w_{\text{in}}$ focused by a thin lens of focal length $f$ is given by $w_0 = \lambda f/(\pi w_{\text{in}})$. In conventional setups, beams with large \(1/e^2\) radius ($w_{\text{in}} > \SI{2.5}{\milli\meter}$) are coupled with long focal length lenses ($f \geq \SI{100}{\milli\meter}$) to achieve narrow waist radii ($\geq \SI{10}{\micro\meter}$) and sufficient trap volumes and depths \cite{GRIMM200095}. However, these laboratory setups typically occupy a large volume of several liters on an optical table and are prone to mechanical and thermal instabilities, which are particularly problematic for crossed-beam ODTs requiring precise and stable alignment of the beam overlap.

The micro-integration approach generally yields systems with high mechanical and thermal stability but limits the apertures of the optical components to obtain a compact footprint, which leads to a smaller input radius of the beam ($w_{\text{in}} \leq \SI{0.5}{\milli\meter}$). Therefore, to achieve a narrow waist, short focal lengths are essential and the positioning of the micro-integrated optical system close to the 3D MOT cell becomes necessary. This also reduces the overall system volume and is beneficial for the overall stability of the physics package. 
In this work, in order to use the shortest possible focal length while maintaining a crossing angle between the two trap beams that is large enough for sufficient atom confinement, the micro-optical bench needs to be inserted within the magnetic field coils of the 3D MOT cell.
Additionally, a clear central aperture through the whole ODT setup is necessary for one of the 3D MOT beams. 
With these requirements and constraints the micro-integrated, crossed-beam optical dipole trap \uxodt~is designed.

\subsection{Optical and mechanical design and system assembly}

\begin{figure}[tb]
    \centering
    \begin{subfigure}[b]{\linewidth}
		\includegraphics[width=\linewidth]{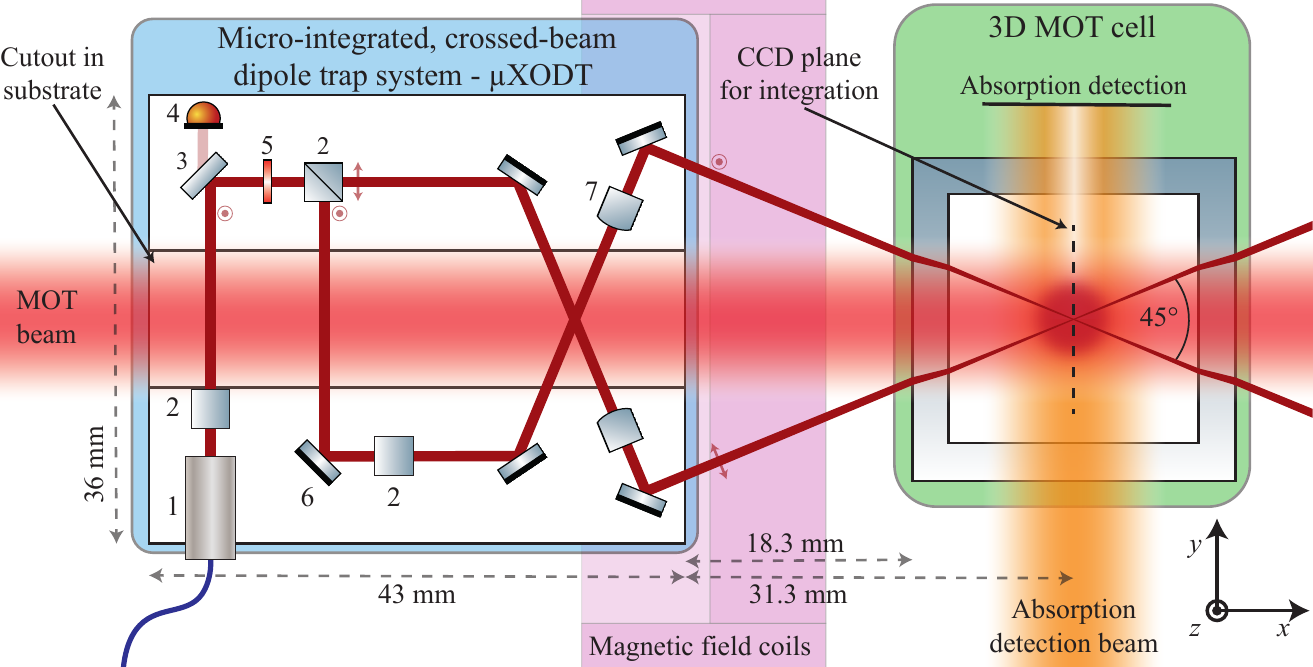}
		\caption{\label{fig:uXODT_opt_layout} Optical layout of the crossed-beam ODT system \uxodt~(light blue box), also depicting the glass vacuum cell of the cold-atom system (light green box) together with the detection axis (shaded in orange). The micro-optical bench with a central cutout to achieve the necessary free aperture for one of the MOT beams (shaded in red) is shown, as well as the magnetic field coils of the cold-atom setup, posing additional dimensional constraints on the optical system indicated in pink. The optical system consists of a collimator (1) with single-mode, polarization-maintaining fiber, PBS (2), partially reflective (3) and highly reflective mirrors (6), a photodiode (4),  $\lambda/2$-waveplate (5), and focusing lenses (7). The orientations of polarization are denoted with arrows pointing inside or out of the plane. The coordinate system used throughout this work is defined as indicated, with gravity pointing along the \textit{-z} direction.
        }
    \end{subfigure}
	\vfill
	\begin{subfigure}[b]{\linewidth}
        \centering
        \includegraphics[width=0.8\linewidth]{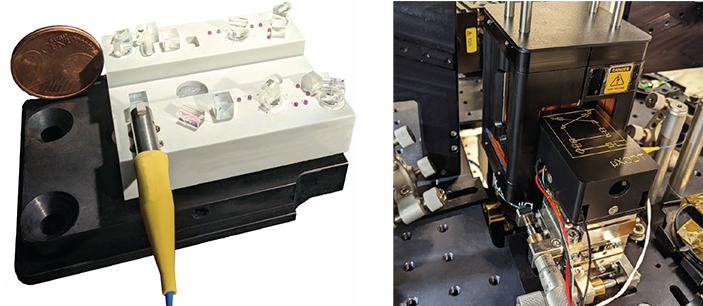}
        \caption{\label{fig:uxodt_trap_depth}Left: the assembled \uxodt~setup mounted on an aluminum adapter plate before integration in the cold-atom system. The optical elements are micro-integrated on a ceramic bench with a central cutout for the MOT beam with a two-cent coin for reference. Right: The \uxodt~system is covered by a lid for protection and integrated with the 3D-MOT setup.}
    \end{subfigure}
    \caption{The optical layout (a) together with pictures (b) of the \uxodt~system and its integration with the cold-atom experiment.}

\end{figure}

The optical layout of the \uxodt~is shown in \cref{fig:uXODT_opt_layout}. The beam shaping optical components to realize the crossed-beam geometry are integrated on a micro-optical bench, while allowing one of the MOT beams to pass through a clear central aperture. The setup is designed to achieve trap depths with a temperature equivalent of about \SI{200}{\micro\kelvin} and angular trap frequencies on the order of $2\pi\cdot\SI{1000}{\hertz}$.
The light to form the dipole trap is coupled from a laser system into the free-space setup via one pigtailed fiber collimator assembly with a single mode, polarization-maintaining fiber. An FC/APC connector facilitates flexible connection to different light sources and integration with the experimental system and laser source described in \cref{sec:BeBEC_uXODT}.

The beam ($1/e^2$ radius of up to $\SI{500}{\micro\metre}$) first passes through a polarizing beam splitter (PBS) and is then partially reflected by a mirror that reflects \SI{99.7}{\percent} of the incident light. The remaining light is transmitted through the mirror via an anti-reflective (AR) coating on the backside and is directed onto a photodiode by an additional mirror (not shown in \cref{fig:uXODT_opt_layout}) for monitoring purposes and power stabilization.
The main part of the light is transmitted through a $\lambda/2$-waveplate and is subsequently separated by a PBS into two paths with perpendicular polarizations to prevent interference in the overlap of both beams. The $\lambda/2$-waveplate allows to balance the intensity between the two beams after passing through the 3D MOT system's glass, accounting for different transmissions of the s- and p-polarized beam paths. An additional PBS cleans the s-polarization in the reflected path, to increase the polarization extinction ratio (PER).

To realize the crossed-beam geometry, each beam is reflected by two mirrors, maintaining reflection angles close to \SI{90}{\degree} (see \cref{fig:uXODT_opt_layout}). This allows the use of standard coatings and reduces the required aperture dimensions. A rectangular plano-convex lens ($f=\SI{40}{\milli\metre}$) is placed between the mirrors for each beam. The beams are precisely overlapped at their waists, crossing at \SI{45}{\degree} angle in the center of the MOT. This crossing point is at \SI[round-mode=places,round-precision=1]{31.25}{\milli\meter} distance from the micro-optical bench, and the bench is placed with \SI[round-mode=places,round-precision=1]{18.25}{\milli\meter} distance from the 3D MOT cell.  

The assembled \uxodt~is shown in \cref{fig:uxodt_trap_depth} with the optical elements micro-integrated on the optical bench. The optical elements have an edge length of \qtyrange{3}{4}{\milli\metre}. For the optical bench, a ceramic material with high specific modulus, low coefficient of thermal expansion (CTE), and high thermal conductivity is selected to achieve the desired mechanical robustness and reduce thermal gradients. It has been shown that these benches can be additively manufactured \cite{Christ2024}. The micro-optical bench also incorporates a Kelvin-type (cone-prism-flat) kinematic coupling to the mounting structure underneath, to reduce mechanical stress from mounting and CTE differences. Furthermore, the substrate features precision alignment and mounting geometries such as stops for the positioning of reference spheres and the optical elements. The required free aperture of \SI{12}{\milli\metre} for the 3D MOT beam is achieved with a central cut-out in the substrate, compare \cref{fig:uXODT_opt_layout}. The bench has a footprint of \qtyproduct[product-units=power]{36 x 43}{\milli\metre}. Including optics, the \uxodt~setup has a mass of \SI{38}{\gram} and an overall volume of approximately \SI{25}{\milli\liter}.

The \uxodt~system is assembled within a dedicated micro-integration facility. Equipped with vacuum tweezers, it allows for the simultaneous alignment of up to four components during the operation of the optical system. Precision actuators can adjust these components in six degrees of freedom with steps as fine as \SI{1}{\nano\meter} and \SI{1}{\micro\radian}. Following the alignment, the micro-optical components are bonded to the substrate using UV, thermal, or hybrid curing adhesives that are specifically chosen for their suitability in the application. Further details on the application and qualification of these adhesives are available in \cite{Christ2023}. 
Micro-optical components such as PBS and some of the mirrors, which are considered less critical, are aligned using precision markers and observing the transmission of the beam path through temporary apertures, placed on the micro-optical bench and within the integration setup. This configuration is expected to achieve an alignment precision finer than \SI{100}{\micro\meter}. The critical alignment steps, i.e., the positioning of the lens and the alignment of the overlap of the two beams, are performed using a bare CCD chip.
This chip is precisely positioned at the designed focal plane relative to the micro-optical bench (compare \cref{fig:uXODT_opt_layout}) and ensures straightforward integration and alignment with the cold-atom system as described in \cref{sec:BeBEC_uXODT}. The use of a bare chip avoids the additional influences of cover glass in the beam path. Continuous image capture allows for the determination of the beam's position and diameter through Gaussian fitting, achieving a positioning resolution smaller than \SI{1}{\micro\meter} for axis alignment and beam overlap. During assembly, a sample glass piece from the 3D MOT cell, featuring the same anti-reflective (AR) coating, is inserted into the beam path to account for its impact on the focal point and transmitted power of the s- and p-polarized beams.

This method minimizes known disturbances in the optical system, ensuring that the focus, overlap quality, and power ratio remains consistent with the original assembly alignment when the \uxodt~is integrated into the cold-atom system. The deformation of the 3D MOT glass cell, induced by bonding and vacuum forces, has not been specifically addressed. However, the deformation magnitude is typically below $\lambda$ peak-to-valley and symmetric along the central \textit{x-z}-plane of the 3D MOT glass cell (refer to \cref{fig:uXODT_opt_layout}), with the maximum deformation in the center of the cell \cite{CQGlass}. Consequently, the impact on both beams is expected to be of minor relevance and the effects are common for both beam paths, thus the effect on the system's performance is expected to be negligible.
The power of both beams is balanced using integrating spheres placed in each beam path and by precisely aligning the \uxodt 's $\lambda/2$-waveplate with the micro-integration setup. After the transmission through the sample cell glass, the relative power ratio of both beams is \num{0.998}.

\subsection{Optical properties and stability evaluation}
\label{sec:breadboard_stability}

\begin{figure}[htb]
	\centering
	\includegraphics[width=\linewidth]{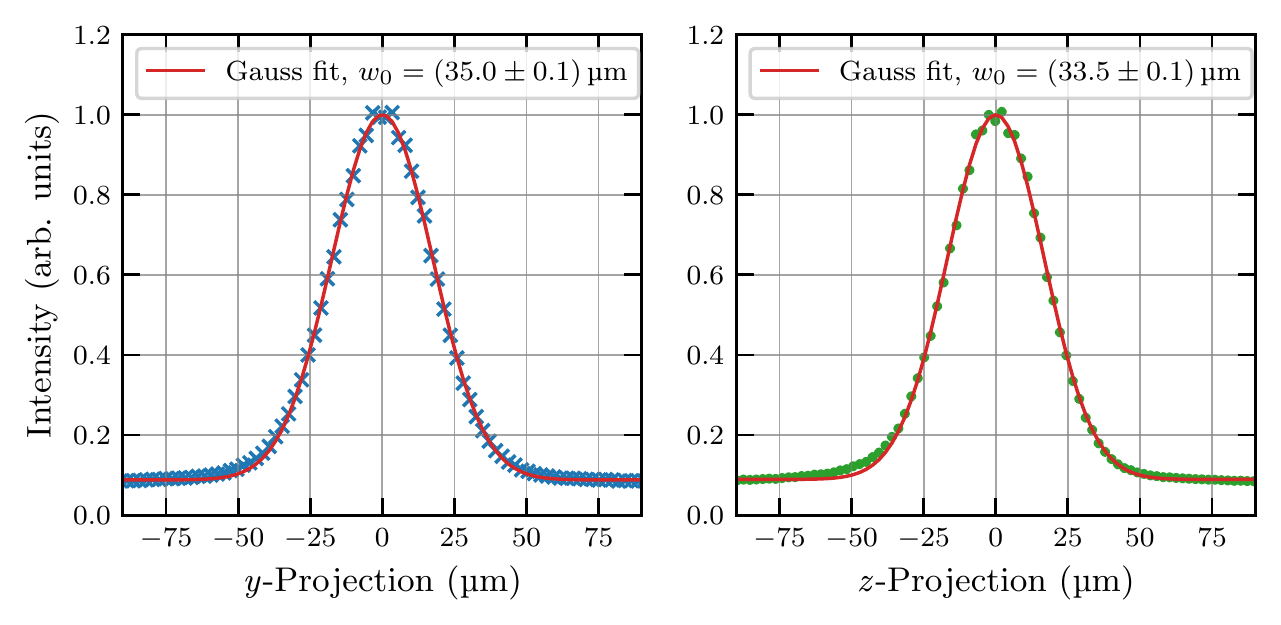}
	\caption{\label{fig:beam_overlap}Intensity profile of the assembled \uxodt. The intensity profile is recorded at the crossing of both beams (compare \cref{fig:uXODT_opt_layout}). The projections follow a single Gaussian profile, indicating good alignment quality, and the $1/e^2$ waist radii of the fit are shown.}
\end{figure}

To characterize the optical properties, overlap quality, and stability of the system, the images captured with the CCD placed in the beam crossing (compare \cref{fig:uXODT_opt_layout}) are evaluated. The \textit{y}- and \textit{z}-projections of the beam profile are depicted in \cref{fig:beam_overlap}. Gaussian fits describe well the recorded data and result in a $1/e^2$ waist radius of \SI{35}{\micro\meter} for the \textit{y}-projection and \SI{33.5}{\micro\meter} for the \textit{z}-projection. 
Due to the orientation of the CCD plane, both beams are incident with an angle of \SI{22.5}{\degree} between the propagation direction and measurement plane, which increases the measured waist radius of the \textit{y}-projection. The measured waist corresponds to $w_0=\SI{32.3}{\micro\meter}$ for a single beam along the propagation direction.

With a typical $M^2=1.1$ for this collimator type, the Rayleigh length is $z_\text{R}=\SI{3.01}{\milli\meter}$ and $z_\text{R}=\SI{2.81}{\milli\meter}$ for the single beam along its propagation direction. The minor ellipticity is already present in the beam emitted from the collimator and can be attributed to manufacturing and assembly tolerances.
This configuration achieves a calculated trap depth of about \SI{200}{\micro\kelvin} for \ce{^{87}Rb} when operated with \SI{2.5}{\watt} total optical power.
All optical components are high power compatible, and the system is assembled in a particle-reduced environment underneath a flow box. The optical system shows no measurable degradation in the quality of the beam overlap and beam position, power ratio, or transmission after several hours with up to \SI{2.5}{\watt} total optical power in continuous wave (CW) operation. We expect the fiber collimator and the fiber mates to be limiting the total optical power. Investigations regarding the maximum power handling of these are ongoing.

\begin{figure}[tb]
	\centering
	\includegraphics[width=\linewidth]{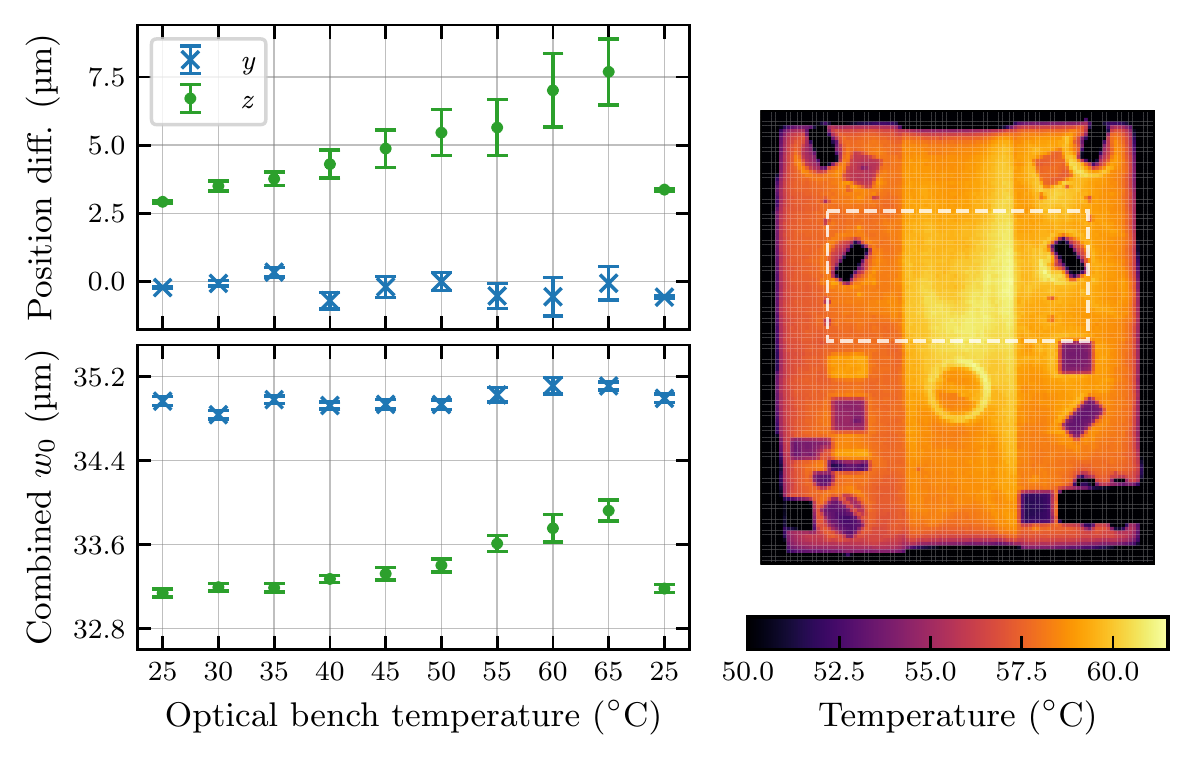}	
	\caption{\label{fig:thermal_stab} Thermal stability of the \uxodt. Left: Position difference between both beams and combined $1/e^2$ waist radius $w_0$ at the crossing during heating and subsequent cool-down of the micro-optical bench (\textit{x}-axis of the plot shows consecutive measurements). Right: thermal profile of the \uxodt~during temperature test recorded with a thermal camera. The position of the thin-film heater, placed underneath the system, is indicated with a white dashed line. The optics appear colder due to the emissivity adjusted for the ceramic optical bench.}
\end{figure}

For thermal stability tests, a resistive thin film heater (\qtyproduct[product-units = power]{12.7 x 25.4}{\milli\metre}, \SI{4}{\watt} heating power) is attached underneath the ceramic micro-optical bench. This allows heating of the micro-optical system up to \SI{65}{\celsius} on an optical table while measuring the beam overlap. During the measurement, the bench temperature is increased from \qtyrange{25}{65}{\celsius} in \SI{5}{\celsius} increments and the CCD is placed in \textit{x}-direction at the beam crossing for each step.
At each temperature increment, 30 consecutive images are recorded for both beams and each beam individually, by blocking the other beam, and a Gaussian fit determines the peak position and beam width. Subsequently, the mean value and standard deviation of the consecutive measurements are calculated. For the single beam measurements, the position difference is calculated as an indicator of overlap quality and stability. The results in \cref{fig:thermal_stab} show a difference in \textit{z}-position between both beams of \SI{2.9}{\micro\meter} at \SI{25}{\celsius}, which can be interpreted as the overall accuracy of the integration including adhesive cure and shrinkage. Taking into account an optical path length of \SI{35}{\milli\meter} from the last mirror to the beam crossing, this equals \SI{83}{\micro\radian} alignment accuracy for this mirror. Further improvements are expected by an optimization of the adhesive joint and a more uniform adhesive dispensing. Detailed studies regarding alignment accuracy, including the bonding process and pointing stability, are currently ongoing. Upon heating, the \textit{z}-position difference increases linearly to \SI{7.7}{\micro\meter} at \SI{65}{\celsius} bench temperature. This effect may be caused by gradients in the micro-optical bench due to non-uniform localized heating or a non-uniform adhesive joint. Although all joints are designed to be symmetric and thus temperature insensitive, small deviations in the adhesive thickness in combination with the high CTE (typically \SI{>50E-6}{\per\kelvin}) can lead to a tilt of the optical component and shift of the beam path. This is especially critical for the collimator, where small alignment deviations of the short focal length lens lead to comparably large deviations of the beam path. 
\Cref{fig:thermal_stab} on the right shows the thermal distribution at \SI{60}{\celsius}, recorded with a thermal imaging camera, along with the position of the resistive heater. The emissivity value is adjusted for ceramics, resulting in an underestimation of the fine-ground or polished glass and metal components on the micro-optical bench. The thermal distribution in \cref{fig:thermal_stab} shows a temperature variation of about \SI{5}{\celsius} over relevant areas of the micro-optical bench. For this measurement the system lid was removed, while it was present during the heating process shown in the plots. With the lid, the temperature is expected to be more homogeneous. The rise in statistical error with temperature is expected to be caused by thermal air convection between the micro-optical bench and the CCD. 
In \cref{fig:thermal_stab} the \textit{y}-position difference is also displayed. This value, however, is only a measure of how accurately the CCD is placed in the crossing and should be close to zero. After holding the system at \SI{65}{\celsius} for two hours and a subsequent cool-down, the \textit{z}-position difference at \SI{25}{\celsius} is \SI{3.4}{\micro\meter}. The minimal deviation from the initial value can be caused by remaining strain in the substrate, changes of the adhesives, water bake-out, or a drift of the camera during the measurement. 

Additionally, \cref{fig:thermal_stab} displays the combined $1/e^2$ waist radius $w_0$ of both beams at the crossing during the temperature profile, showing a constant $1/e^2$ waist radius for the \textit{y}-projection. The waist radius in \textit{z}-direction increases with increasing temperature, caused by the position difference of both beams. When heated from \qtyrange{25}{65}{\celsius}, the $1/e^2$ waist radius increases by less than \SI{1}{\micro\meter} (\SI{3}{\percent}). After cooling down to \SI{25}{\celsius}, $w_0$ returns to the initial value for both axes.

Since assembly, the \uxodt~system has been transported to multiple labs, stored in different locations, and integrated into various setups for micro-integration and measurement. Furthermore, the system has been integrated with a cold-atom experiment (see \cref{sec:BeBEC_uXODT}) and operated for over two years. No substantial changes in the optical properties of the system have been measured so far. This underlines that high thermal and mechanical stability is a key advantage of miniaturized, micro-integrated optical systems, in addition to their optimized size and weight.

\section{Operation of the \textmu XODT in the cold-atom experiment}
\label{sec:BeBEC_uXODT}

To evaluate the \uxodt~for atom trapping, the system is integrated into the cold-atom system as described in \cref{sec:uxodt_design_assembly_tests}, and shown in \cref{fig:uxodt_trap_depth}. For protection, the system is covered by an aluminum lid. The MOT beam passes through an aperture in the lid and then through the substrate cutout. Electrical connections for the \uxodt~are available for temperature measurement, connecting the photodiode and a Peltier element. The latter was included for potential temperature stabilization, which proved to be not necessary. To compensate for differences in the coefficient of thermal expansion and minimize induced mechanical stress, the micro-optical bench rests on a kinematic mounting structure as introduced in \cref{sec:uxodt_design_assembly_tests}. Since the position of the 3D MOT cell is not precisely determined with respect to the breadboard, the \uxodt~is mounted on top of a compact opto-mechanical stack of precision actuators, allowing for alignment in the relevant positions \textit{x} and \textit{y}, pitch \textit{Ry}, and yaw \textit{Rz} axis. With the linear axes, the crossing position can be moved with respect to the center of the MOT cell. The rotations allow for an alignment of the beam path with respect to the glass cell. 
In addition, the position of the 3D MOT can be moved within the glass cell using the offset magnetic fields generated by the coils.

\subsection{Laser source for the µXODT}

\begin{figure}[tb]
\centering
\includegraphics[width=\linewidth]{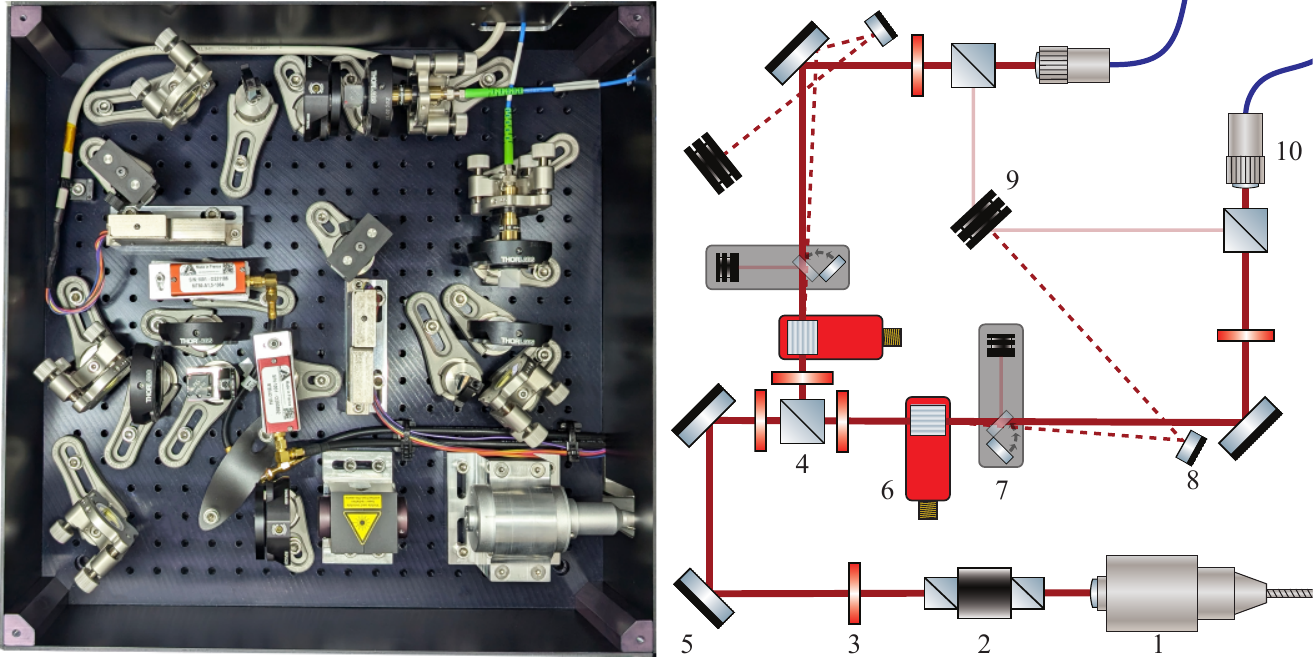}
\caption{\label{fig:laser_system} Transportable laser distribution and modulation system for operation of the \uxodt. Left: photograph of the system with a \qtyproduct[product-units=power]{410 x 410}{\milli\metre} footprint and two fiber coupled outputs. Right: optical layout of the system, consisting of: output collimator of the fiber amplifier with optical isolator (1), additional optical isolator (2), $\lambda/2$-waveplates (3), PBS (4), highly reflective mirrors (5), AOM (6) with $0^{\text{th}}$ (dashed line) and $1^{\text{th}}$ diffraction order (solid line), electrically actuated optical shutters with internal flip-mirror and beam dump (7), highly reflective D-mirrors (8), beam dumps (9), and fiber couplers with adjustable lens and FC/APC connector (10).}
\end{figure}

The laser system for operating the \uxodt~consists of a high-power light source and an optical system designed for power distribution, modulation, and stabilization. The light source is composed of a single-frequency, narrow linewidth seed laser (NKT Photonics Koheras ADJUSTIK Y10) operating at \SI{1064}{\nano\metre}, and a fiber amplifier (NKT Photonics Koheras Boostik HP Y10) capable of delivering up to \SI{5}{\watt} of continuous wave output power. The amplifier output collimator is integrated into a free-space optical system housed in an anodized aluminum box (volume \qtyproduct[product-units=power]{410 x 410 x 110}{\milli\metre}, \SI{1}{inch} standard optics). A photograph of the system and a schematic of the optical layout are displayed in \cref{fig:laser_system}. The light is distributed into two paths with an adjustable power ratio, which are then coupled into polarization-maintaining fibers. Each path includes an electrically actuated mechanical shutter, an acousto-optic modulator (AOM), $\lambda/2$-waveplates and an additional polarizer with an extinction ratio greater than $\SI{35}{\deci\bel}$. The AOM facilitates optical power stabilization and the implementation of ramps and pulse shaping for applications such as evaporative cooling or optical lensing. Power stabilization is implemented using a photodiode and a PID controller running on an FPGA (Red Pitaya STEMlab 124-14), achieving a relative power instability well below $10^{-4}$ for experimental cycle times of up to \SI{100}{\second}. Initially, all light for the operation of the \uxodt~is directed into one fiber. However, the second path is available for future applications, as discussed in \cref{sec:outlook}. The seed laser and fiber amplifier in \SI{19}{inch} format and free-space system in a dedicated housing allow for transportation of the laser system to different laboratories.

\subsection{Dipole trap lifetime and stability}
\begin{figure}[tb]
\centering
\includegraphics[width=0.8\linewidth]{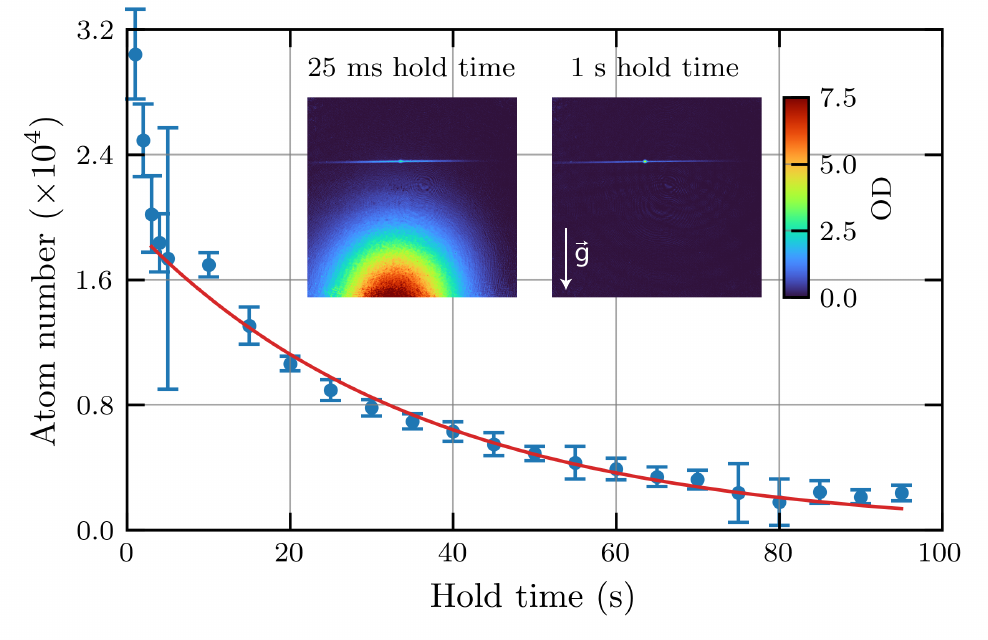}
\caption{\label{fig:bebec_lifetime}\label{fig:bebec_absorption} Trap lifetime measurement. \ce{^{87}Rb} atom number in the dipole trap operated with \SI{\approx 1.5}{\watt} optical power for different hold times after switching off the MOT. The trap is loaded from the released atoms and the trap wings for hold times up to \SI{2}{\second}. An exponential fit (\(y(t)=Ae^{-t/\tau}\), plotted in red) of the data for hold times \SI{>2}{\second} yields a lifetime of \lifetime. The inset shows absorption images of the ODT, taken after \SI{25}{\milli\second}  and \SI{1}{\second} hold time.}
\end{figure}
The experimental sequence to load atoms into the dipole trap starts with loading the MOT for \SI{2}{\second}, followed by a compression and optical molasses stage for \SI{25}{\milli\second}. This yields approximately \num{1e8} atoms at a temperature of \SI{6}{\micro\kelvin} \cite{Anton2024}. At the end of the MOT loading, the dipole trap is switched on to an optical power of \SI{730}{\milli\watt} per beam.
The light of the MOT is switched off and the atoms are held in the ODT for at least \SI{100}{\milli\second}. During this time, atoms that are not transferred into the ODT undergo free fall under gravity and exit the camera's field of view, leaving only atoms visible that are trapped in the wings and center of the ODT. This allows for counting of the number of atoms inside the ODT without the background interference from the cloud.
With this sequence, up to \num{4e5} atoms are loaded into the dipole trap. The number of atoms is determined from absorption images using light resonant with the \(\lvert F=2 \rangle \rightarrow \lvert F'=3 \rangle\) transition, together with the position of ODT in the images. The orientation of the absorption detection with respect to the crossing beams of the ODT is indicated in \cref{fig:uXODT_opt_layout}.
The inset of \cref{fig:bebec_absorption} shows two selected absorption images after holding the atoms for \SI{25}{\milli\second} and \SI{1}{\second}, with the molasses visible in free fall below the ODT and with transferred atoms remaining in the ODT and its wings, respectively.
 
The lifetime of atoms trapped in the ODT is determined to characterize the trap. 
\Cref{fig:bebec_absorption} displays the number of atoms inside the ODT for hold times ranging from \qtyrange{0}{95}{\second} after switching off the MOT light fields. 
A step size of \SI{1}{\second} is used for hold times from \qtyrange{1}{5}{\second} with \num{10} repetitions each step. For times from \qtyrange{10}{95}{\second}, \num{5} repetitions are performed for each \SI{5}{\second} step. We determine the standard deviation of these measurements, which are given as error bars.
From the recorded absorption images, we determine that the dipole trap is loading from its wings up to a hold time of \SI{2}{\second}. An exponential fit (\(y(t)=Ae^{-t/\tau}\)) for times \SI{>2}{\second} yields a lifetime of \lifetime, assuming single-particle losses as limiting mechanism for these timescales \cite{Kuppens2000}. This lifetime is suitable for applications based on cold atoms.

\begin{figure}[tb]
	\centering
	
        \includegraphics[width=\linewidth]{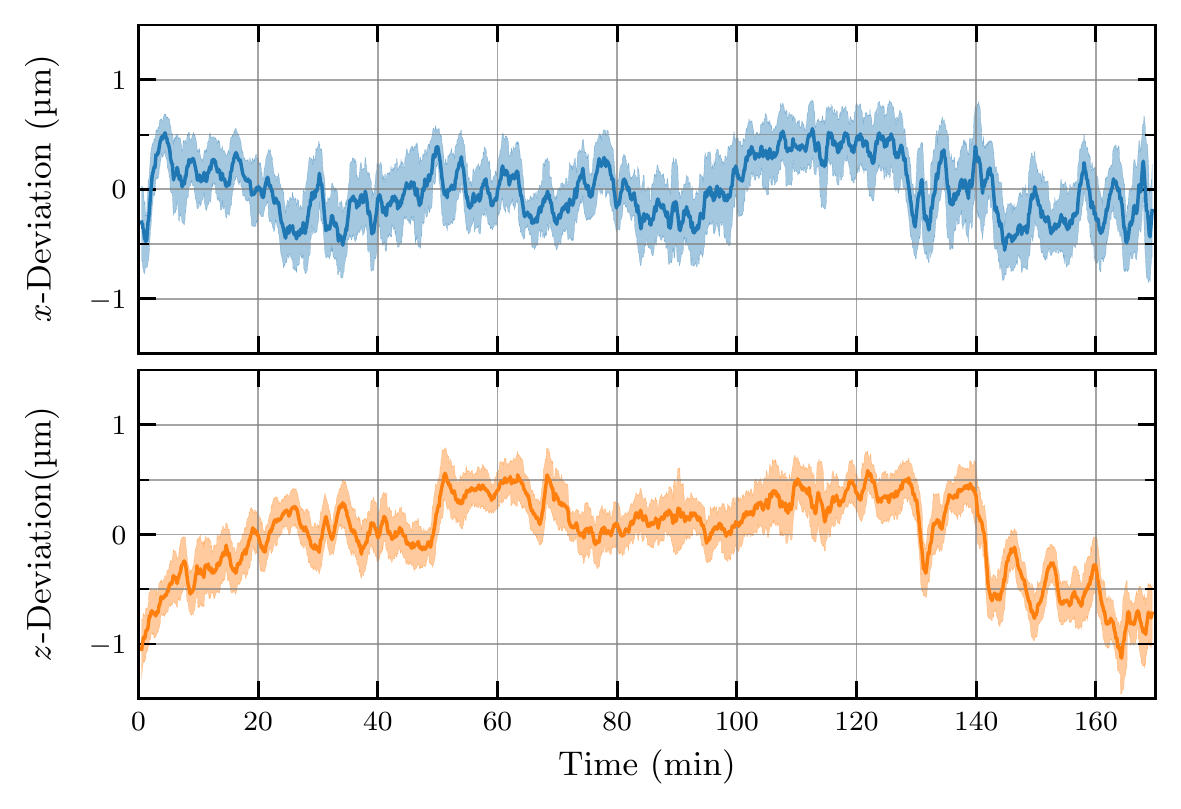}
	
	\caption{\label{fig:bebec_stability} Temporal stability of the dipole trap position. The \textit{x}- and \textit{z}-position of the ODT is determined from 3000 consecutive absorption images, each taken after an experimental cycle time of \SI{3.4}{\second} including \SI{2}{\second} hold time, and the deviation from the mean value calculated. The solid line indicates a rolling average of \num{20} measurements, and the standard deviation is shown as a shaded area. The atom number during this measurement series is \num[scientific-notation=true, round-mode=uncertainty, round-precision=1, separate-uncertainty=true]{330810 \pm 18897}.}
\end{figure}

The stability of the trap position in the experimental system is evaluated from \num{3000} consecutive experimental cycles, each lasting \SI{3.4}{\second} with \SI{2}{\second} hold time. From the recorded \textit{x}- and \textit{z}-positions of the ODT, the deviation from the mean value is calculated and displayed in \cref{fig:bebec_stability}. Across the performed measurement series, the standard deviation of the \uxodt 's position on the CCD is less than \SI{\pm 0.5}{\micro\meter}. This deviation is likely due to thermal effects on the experiment's opto-mechanical components, resulting in a relative drift of the \uxodt's micro-optical bench or the mount, but does not affect the beam overlap through the micro-integration approach (compare \cref{sec:breadboard_stability}). This represents a significant improvement compared to traditional breadboard ODT systems, where the overlap would typically suffer from drifts. The atom number recorded in these measurements is \num[scientific-notation=true, round-mode=uncertainty, round-precision=1, separate-uncertainty=true]{330810 \pm 18897}. Over the course of two years, the stability and ODT properties remain on this level, indicating no degradation of the dipole trap system.

\section{Summary and outlook}
\label{sec:outlook}

In this work, the development and application of a micro-integrated optical system for cold-atom applications is presented. By adapting the integration approach from miniaturized diode lasers, a crossed-beam optical dipole trap setup, the \uxodt, is realized.
The fiber-coupled free-space setup is realized by micro-integrated optical components on an optical bench of \qtyproduct[product-units=power]{36 x 43}{\milli\metre} footprint. It generates two focused beams, overlapping in their focal point with a crossing angle of \SI{45}{\degree}.
The volume of approximately \SI{25}{\milli\liter} and mass of \SI{38}{\gram} including optical components represent a substantial reduction compared to lab-based setups that typically employ macroscopic opto-mechanical components and occupy volumes of several liters. The system demonstrates long-term alignment stability after thermal heating to \SI{65}{\celsius}, CW operation with optical powers up to \SI{2.5}{\watt}, transport to different laboratory facilities, and integration of the \uxodt~into a \ce{^{87}Rb} 3D MOT system and operation for more than two years. We load \num[scientific-notation=true, round-mode=uncertainty, round-precision=1, separate-uncertainty=true]{330810 \pm 18897} atoms from the molasses, with an atom lifetime of \lifetime.

Overall, the measured lifetime and atom numbers are suitable for a broad range of cold-atom based applications. The compact volume, high stability, and reliability of the dipole trap system highlight the benefits of micro-integrated atom traps for compact physics packages. This becomes especially relevant for mobile applications, where the fiber-coupled setup can be integrated into higher-level cold-atom systems. The presented dipole trap complements parallel efforts towards compact and simplified physics packages, such as next generation atom chips \cite{Kassner2019, Heine2024} or grating-based MOTs \cite{Cotter2016}.

\begin{figure}[tb]
	\centering
	\includegraphics[width=\linewidth]{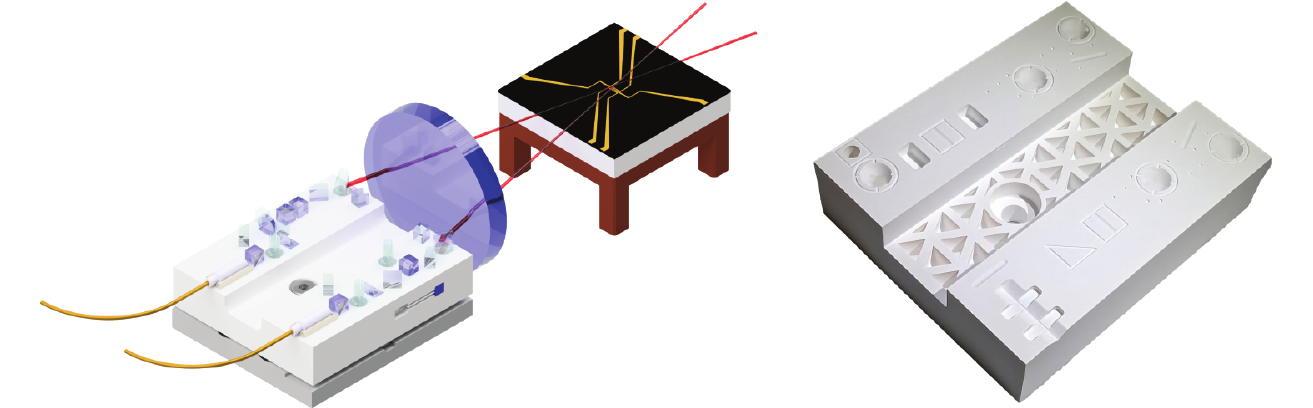}
	
	\caption{\label{outlook}Left: CAD rendering of the extended ODT for manipulation of ultra-cold atoms in an atom-chip based setup, including the vacuum window and an exemplary atom chip. The extended trap has a footprint of \qtyproduct[product-units=power]{42 x 48}{\milli\metre} and will realize a \SI{25}{\degree} crossed beam geometry trough a single vacuum viewport with \SI{27}{\milli\metre} aperture. Right: additively manufactured micro-optical bench for a crossed beam ODT. The ceramic substrate is shown after printing and before post-processing.}
\end{figure}

Current developments focus on integrating a micro-integrated, crossed-beam ODT system within an ultra-cold-atom setup employing an atom chip \cite{Christ2019}. Here, the ODT can enhance the weak magnetic trapping axis of the atom chip for enhanced evaporative cooling. First, a similar approach of placing the optical setup in front of the vacuum system is pursued. The design of the extended dipole trap setup is displayed in \cref{outlook} and is to be combined with next generation atom chips \cite{Kassner2019, Heine2024}. The vacuum chamber size and available viewport aperture of \SI{27}{\milli\meter} restrict the dipole trap to crossing angles of $\SI{25}{\degree}$ and makes lenses of $f\approx\SI{70}{\milli\meter}$ necessary. This increases the $w_0$ of the dipole trap and reduces the trap potential and frequencies.
A subsequent development integrates a micro-integrated ODT directly with the atom chip holder within the vacuum system. This allows for more favorable trapping geometries through short focal lengths ($f\leq\SI{30}{\milli\meter}$) and large crossing angles ($\approx\SI{70}{\degree}$) that are otherwise inaccessible given by the constraints imposed by the vacuum chamber. Furthermore, a rigid and well-defined connection with the atom chip is essential to omit the opto-mechanical stages and thus further reduce the overall SWaP requirements. However, this new approach requires ultra-low outgassing components, adhesives qualified for this specific application, and an individual adaptation of the trap to each fabricated chip system, which is necessary for precise alignment to the center of the chip's magnetic trap. R\&D efforts on these topics are currently ongoing \cite{Christ2023}.

Future work includes increasing the functionalization of micro-integrated ODTs. The micro-optical bench can be printed from ceramics, which is a material of choice due to the high specific modulus, suitable CTE, and sufficient thermal properties \cite{Christ2024}. A printed micro-optical bench is displayed in \cref{outlook}. This approach enables functionalized designs while minimizing the size and weight through optimization. Additionally, the implementation of active optics for beam steering is being investigated, for example with MEMS mirrors or compact acousto-optic deflectors. This would enable the implementation of painted potentials for an enhancement of the trap volume, among other applications \cite{Barredo2016, Bregazzi2023}.

\section*{Author contributions}
M.C. developed and designed the \uxodt, the assembly concept and the integration with the cold-atom system, micro-integrated the \uxodt~and conducted the optical qualification, evaluated the measurements presented in this work, and drafted the manuscript; M.C. and O.A. integrated the \uxodt~with the cold-atom system; O.A. and E.D.R. operated the cold-atom system, performed the presented lifetime measurements and evaluated the absorption images with support from V.H.; C.Z. inspected the micro-optical components, wrote the software for CCD readout and supported the micro-integration; M.C. and C.Z. build the high-power laser system; M.C. and M.K. acquired funding and conceptualized the work. M.K. supervised the work. All authors read and reviewed the manuscript. M.C. revised the manuscript.

\section*{Acknowledgements}
The authors thank Anne Stiekel and Alisa Ukhanova for their support in building and assembly of the high-power laser system. The authors thank Simon Kanthak for support in ODT property calculation, technical advice on the high power laser system design, and critical review of the manuscript. M.C. and C.Z. thank Max Schiemangk, Robert Smol, and Christian Kürbis for technical advice regarding micro-integration and Sascha Neinert for processing the printed ceramic micro-optical bench. The authors thank Matthias Schoch for technical support on electronics.

This work is supported by the German Space Agency (DLR) with funds provided by the Federal Ministry for Economic Affairs and Climate Action (BMWK) due to an enactment of the German Bundestag under grant number 50WM1949 (KACTUS-II), 50RK1978 (QCHIP), 50WM2070 (CAPTAIN-QT), and 50WM2347 (OPTIMO-III).

\bibliography{references}

\end{document}